# A Service Suite for Specifying Digital Twins for Industry 5.0

Izaque Esteves[1], Regina Braga[1,*] orcid: 0000-0002-4888-0778, José Maria N. David[1] orcid: 0000-0002-3378-015X, Victor Stroele[1] orcid:0000-0001-6296-8605

izaque.esteves@estudante.ufjf.br, {regina.braga, jose.david, victor.stroele}@ufjf.br

[1]Computer Science PostGraduate Program
Federal University of Juiz de Fora, MG, Brazil.

 * **Corresponding author:** Regina Braga – regina.braga@ufjf.br

**Abstract**  One of the challenges of predictive maintenance is making decisions based on data in an agile and assertive way. Connected sensors and operational data favor intelligent processing techniques to enrich information and enable decision-making. Digital Twins (DTs) can be used to process information and support decision-making. DTs are a real-time representation of physical machines and generate data that predictive maintenance can use to make assertive and quick decisions. The main contribution of this work is the specification of a suite of services for specifying DTs, called DT-Create, focused on decision support in predictive maintenance. DT-Create suite is based on intelligent techniques, semantic data processing, and self-adaptation. This suite was developed using the Design Science Research (DSR) methodology through two development cycles and evaluated through case studies. The results demonstrate the feasibility of using DT-Create in specifying DTs considering the following aspects: (i) collection, storage, and intelligent processing of data generated by sensors, (ii) enrichment of information through machine learning and ontologies, (iii) use of intelligent techniques to select predictive models that adhere to the available data set, and (iv) decision support and self-adaptation.

**Keywords**: *intelligent techniques, internet of things, sustainability, predictive maintenance, decision support, digital twins, self-adaptive systems, industry 5.0,*

## 1      INTRODUCTION

The current perception of industry results from transformations driven by the need for high-quality production and faster delivery capacity. The consolidated stage of industrial development is called Industry 4.0, characterized by using advanced technologies and the Internet (Lasi et al., 2014). In this stage, we have "things" (equipment) that can process data and establish connections with other devices through communication networks, which characterize the Internet of Things

(IoT) (Ashton, 2009), an essential component in Industry 4.0. However, we are in a transition to a new stage: Industry 5.0. According to the European Commission for Research and Innovation (Breque et al., 2021), Industry 5.0 is an extension of Industry 4.0, which has humans as the central actors in the production process, in addition to focusing on the sustainability of the industry based on the increasingly scarce resources and on resilience. The aim is to create a value chain that goes beyond cost-efficiency or profit maximization but benefits all stakeholders: investors, workers, consumers, society, and the environment (Breque et al., 2021). Observing the enabling technologies of Industry 4.0 and the objectives of Industry 5.0, one of the challenges related to industrial maintenance processes is the viability of the transition from corrective and preventive maintenance models to predictive maintenance. Predictive maintenance anticipates problems based on data collection and preprocessing, defining a model for data analysis, which enables the creation of mechanisms to assist in decision-making (Achouch et al., 2022).

A predictive maintenance life cycle begins with an understanding of the problems and constraints of a project, establishing the quantities to be measured, which sensors will be installed and where, and identifying possible types of failures. With the infrastructure in place, data collection is performed by sensors on the equipment. The understanding and preparation stage aims to identify which data will be analyzed and to construct meaning for the data. Thus, the data goes through cleaning processes, data volume management, corrections of missing or anomalous data, and selection of data types, among other processes. In the data modeling stage, the most appropriate procedure is chosen so as to create a model capable of making predictions with the best possible accuracy. The model is tested for its accuracy and relevance in the implementation stage. The cycle ends with decision-making, which includes solving problems and making decisions, selecting the most appropriate procedures to minimize costs and delays, and evaluating decisions to improve future interventions. Analyzing this life cycle, we can see that involved industrial processes increasingly focus on anticipating decisions more automatically. In addition, there is a concern with sustainability since the processes focus on the longevity of equipment. In this sense, predictive maintenance can be considered an essential process for Industry 5.0.

However, one of the main challenges is the lack of data on events of interest as the machine is often repaired before any typical event for maintenance occurs. To establish a virtuous cycle, we need specific techniques to capture data on events, without interventions in the production system. Digital Twins (DTs) provide a real-time representation of the physical machine and generate data that can be used in predictive maintenance (van Dinter et al., 2022). Thus, predictive maintenance can be implemented without directly affecting the equipment in production. DTs are digital replicas of a physical system or component which use data from the physical system as inputs to simulate how this data affects the production system.

However, even with DTs, the maintenance scenario is challenging, considering the speed at which decision-making must occur. Additionally, collaboration among stakeholders is necessary in decision-making processes, as it takes advantage of expertise from different areas and positions. The agility in producing information also gives decisions a greater degree of anticipation regarding failures and assertiveness. It is therefore necessary to create a decision-making scenario which is rich in information and uses technologies that enable the early discovery of information in an agile manner.

As a result, interest in new software development strategies is growing in search of new approaches for developing complex systems (Weyns, 2020). Self-adaptive software engineering (Eramo et al., 2021), encompassing context awareness and intelligent strategies, is an approach that aims to solve complexities from a new perspective. The Industry 5.0 maintenance scenario requires speed in decision-making while also considering human factors and sustainability consistently. One of the difficulties is related to real-time monitoring and treatment of operational problems in production processes. In addition, there are challenges involving resource management, increasingly customized production, and processes that require management with increasingly faster responses. In this context, it is necessary to use technologies that favor the development of intelligent solutions for these demands. Self-adaptive architecture can help in this complex scenario.

In addition, Machine Learning (ML) techniques are inserted in an environment where predictive models are selected according to the dataset, enabling the discovery of knowledge about data with greater assertiveness. In a complementary way, semantic models, such as ontologies (Gruber, 1995), enrich the information obtained from the evaluation of predictive models, additionally to important contextual information for decision-making. We, therefore, combine syntactic and semantic analyses for data enrichment (Amershi et al., 2019) and self-adaptive strategies to take specific actions in the production system.

To address these challenges, this work is guided by the following Research Question (RQ): **"How can decision-making related to predictive maintenance be supported in industry using DTs?"**. We employed the Design Science Research - DSR (Hevner et al., 2008) approach to underpin the research through the artifact produced. DSR is an approach characterized by the continuous improvement of a solution through the introduction of new artifacts and processes for building them (Simon, 1996). DSR involves the creation of models, prototypes, or systems and their evaluation in terms of usefulness, effectiveness, and impact. DSR also involves evaluation cycles. We conducted 2 evaluation cycles, in real scenarios. The first in a textile industry and the second one with data from sensors installed in heat treatment furnaces.

The paper is divided into six sections. Section 2 presents related works. Section 3 presents the methodology used and the suite of services to support decision-making in predictive maintenance, detailing the conceptual aspects and the implementation. Section 4 details two evaluation cycles, highlighting the planning, execution, and results obtained. Section 5 presents the final considerations and discusses the contributions of the work, its limitations, and future work.

## 2    Related Works

An evidence-based mapping study uses a systematic and transparent approach to identify, evaluate, and synthesize the best evidence on a specific topic. It is a comprehensive and rigorous study of all available evidence on a given topic, conducted according to pre-specified criteria (Kitchenham and Brereton, 2013). In order to verify how literature discusses the subject, we conducted a systematic mapping considering scientific articles that discuss data analysis from IoT devices, processed by intelligent techniques, in the predictive maintenance context.

We adopted the hybrid approach proposed by Mourão et al. (2020). The hybrid strategy combines searches in digital libraries through backward snowballing (BS) and forward snowballing (FS). The searches were carried out in the Scopus digital library. The comparative analysis with other digital libraries (Mourão et al., 2020) offered better results when the protocol used combined the search with SB and SF and was compared with other digital libraries. This mapping study is available at [1].

On analyzing the set of returned papers, it was found that many papers consider preventive maintenance as a key issue in Industry 4.0 systems. However, these papers do not consider the complexity of software systems required by Industry 4.0. Rojek et al. (2023) present a solution to implement a maintenance strategy based on the historical analysis of failures to classify them according to their types, causes, and maintenance activities. The solution also offers a model for predicting failures. It does not advance in supporting preventive maintenance decisions as it analyzes historical data, nor does it offer self-adaptation and management resources, nor indicate the need for interventions.

Ayvaz and Alpay (2021) also present a solution for predicting failures, but it delves deeper into the selection of a machine learning model with greater predictive capacity. The solution addresses the selection of the most important features in the dataset but does not select a predictive model based on these characteristics, and this selection is not done automatically, i.e., with each change in the dataset, a new predicted model must be chosen according to its accuracy.

Ortiz et al. (2019) present a solution based on microservices. The contextual approach that the solution proposes is noteworthy, considering that in general IoT systems are context-sensitive, offering a context-sensitive and adaptable predictive analysis system and different application scenarios. However, the proposed solution does not allow the selection of the most appropriate predictive model, and the data produced by the predictive model is applied only to alert users but is not used for self-adjustment of the evaluated system.

Chen et al. (2021) propose a data-driven predictive maintenance strategy for maintenance decision-making for repairable engineering systems. The strategy only considers the LSTM

---

[1] Available at https://repositorio.ufjf.br/jspui/handle/ufjf/16844 (in Portuguese)

algorithm to analyze the data set and make predictions. There is also no enrichment of information about the system's state from knowledge extraction, and there is no interface between the predictive model and the current system beyond maintenance prediction, such as triggering alerts for upcoming maintenance or emergency intervention in some equipment.

Peres et al. (2018) present a framework for intelligent data analysis and real-time supervision, combining distributed data acquisition, machine learning, and runtime reasoning. This framework contributes to predictive maintenance and quality control and reduces the impacts of disruptive events on production. The proposal aligns with our proposal but does not explore the potential that information enrichment and self-adaptation provide.

Finally, Neto et al. (2021) propose a DT for scheduling opportunistic predictive maintenance. The DT represents the stock of products during the manufacturing process, with information obtained through RFID and the operational status of the machines obtained through sensor data. Decision support is provided through periodic simulations. The proposal contributes to maintenance only in its planning without offering technology for producing knowledge from simulations. Also, machine learning is not applied to selecting which model is most adherent to the current production system model, thus increasing the model's accuracy in finding maintenance opportunities.

Our work uses machine learning techniques in the context of large datasets and supports the predictive maintenance process through a DT, also enabling the selection of the best predictive model for a given dataset. Unlike the works returned in the systematic mapping, our proposal enriches information from ontology models and inferences to assist in decision-making regarding industrial processes, providing the system with the ability to be self-adaptive given that the system itself uses the intelligent selection of predictive models to make the system self-adaptive with the most assertive model for the dataset.

## 3   DT-CREATE – SERVICE SUITE FOR DIGITAL TWINS SPECIFICATION IN INDUSTRY 5.0

Based on the mapping study results, we identified gaps in predictive maintenance support, mainly related to data analysis using machine learning models. In addition, none of the selected works uses semantic models combined with artificial intelligence algorithms to favor self-adaptation. Considering these gaps, we developed DT-Create, a suite of services supporting data analysis in industry, aiding decision-making in predictive maintenance.

### 3.1   **Methodology**

Design Science Research (DSR) is an approach to building solutions by elaborating, developing, and evaluating artifacts. Knowledge and understanding of a problem domain and its solution are achieved by constructing and applying the designed artifact (Hevner et al., 2007). The method aims to develop solutions to real problems, combining scientific rigor with practical relevance. The construction of artifacts follows scientific methods and aims to generate knowledge about a specific item. The artifact design corresponds to an iterative and incremental activity, and its evaluation occurs at each DSR cycle, providing feedback for building and improving the product (Hevner et al., 2008).

We conducted the following steps in the DSR: problem definition, literature review, discussion of existing solutions, artifact development, evaluation, and discussion of results. In the first step, we identified the Research Question (RQ): **"How can decision-making related to predictive maintenance be supported in industry using DTs?"**. As a next step, we investigated related works to find proposals in the literature that deal with decision support in industry. To this end, we conducted a systematic mapping, reported in Section 3. Based on the findings, we observed gaps in the studies that reveal the need to explore solutions capable of dealing with the complexity and diversity of data and possible solutions in the industrial context. We consider that by aggregating the results of the use of intelligent techniques with semantic models focused on predictive maintenance, decision-making processes regarding maintenance can be improved, considering both support for decision-makers and issues related to sustainability. We therefore developed a suite of services in two DSR cycles. In each cycle, we conducted evaluations through real-world case studies, which generated scientific and practical knowledge.

In the initial development of DT-Create, we identified the key functional (FR) and non-functional requirements (NFR) that are priorities for its progress. The functional requirements include: (i) Process different sets of industrial data on maintenance or even operation (FR01); (ii) Integrate context information, to help enrich the dataset (FR02); (iii) Perform semantic and machine learning analysis (FR03); (iv) Allow efficient data storage (FR04); (v) Provide mechanisms for visualizing processed data to support decision-makers in interpreting information and making decisions (FR05); (vi) Support the essential requirements of Industry 5.0-related solutions (FR06). As for non-functional requirements, we can highlight the following: (i) The solution must enable reliable communication with data sources (NFR01); (ii) it should allow for extensibility to encompass new functionalities (NFR02); and it needs to evolve according to the context of new datasets (NFR03). Flexibility is a key attribute, reflecting the range of behaviors the existing solution must address, such as processing machine operation data, maintenance processes, and other data related to industrial product processes.

### 3.2 First DSR Cycle

Based on the results of systematic mapping and functional and non-functional requirements, it was possible to establish the conjectures for the first version of the suite of services (Da Silva, et al., 2021). This first cycle led to the construction of an artifact with functionalities for data analysis combined with ontological processing to provide decision support information. From the DT-CREATE version derived from the first DSR cycle, it was possible to specify a DT to support predictive maintenance in a textile industry. We conducted an evaluation and identified aspects for improvement in DT-Create and the derived DT. More specifically, we identified the following limitations: (1) According to the nature of the dataset analyzed, the use of a single type of machine learning algorithm could make the process less effective since we do not have a learning model that is efficient in its predictive capacity in any context, that is, with any dataset. (2) Self-adaptation proved to be a possibility for improvement, considering that, based on the input data, it is possible to select the most appropriate intelligent processing technique. Furthermore, with the evolution of equipment and solutions in Industry 5.0, it is possible to make adaptations in the production system, such as shutting down equipment and redirecting production to another production system, among others. Therefore, we identified an opportunity to evolve DT-Create to adopt new self-adaptation features and the possibility of selecting intelligent data processing, improving the creation of DTs from DT-Create.

Therefore, we added the following functional requirements. As functional requirements, we used different intelligent processing techniques, according to the input data (RF07), and followed new functionalities related to self-adaptation (RF08). So, we executed an additional DSR cycle.

### 3.3 Second DSR Cycle

This cycle allows the selection of a predictive model from a set of available models that are more assertive considering a given input dataset. In addition, the suite of services incorporated new self-adaptive features. The intelligent component was improved and is called AutoML, according to Figure 1, reflecting the changes related to the selection of ML models. In the second DSR cycle, DT-Create encompasses a process of training and evaluating a set of predictive models, with the automatic selection of the best model available for use with the input dataset.

The new AutoML component executes some steps based on the input dataset. The first step is the selection of the relevant attributes, which consists of selecting features in the dataset that contribute to predicting the target variable, thereby reducing the risk of over-fitting, improving its accuracy and reducing training time. The second step, training multiple models, aims to find the model that best fits the dataset. In this step, a function trains and evaluates the performance of selected models using cross-validation. Based on the average cross-validation scores, the most suitable mode is selected, which undergo hyperparameter adjustments (tuning) to increase its predictive capacity. Then, the chosen predictive model becomes the reference model for evaluating new data.

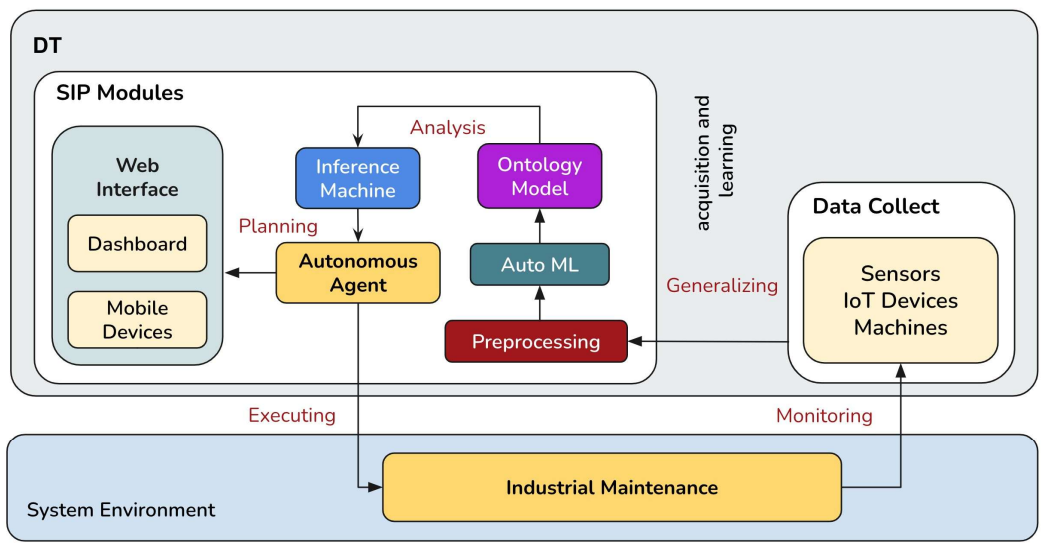

Figure 1 - DT-Create architecture - second DSR cycle.

Figure 1 presents a conceptual model of DT-Create and its main components:

• **System Environment**: Represents the physical space where the production devices are installed and, through a data communication network infrastructure (Message Queuing Telemetry Transport - MQTT[2]), telemetry data is transmitted. Older machines generally do not support connectivity in industrial networks but can be connected through IoT devices. Thus, monitoring the environment consists of capturing telemetry data from IoT and sensor data, directing them to the Data Collect component.

• **Data Collect**: Offers data integration from various devices through the MQTT communication protocol, allowing status information to be sent to a central device. Devices are connected to a central node called a Broker, which adopts the communication model oriented towards publishing and subscribing to topics. The Broker acts as a dispatcher for IoT devices and other connected equipment storing the data collected on a server. The data is used by the AutoML component (Figure 1), passing through the Preprocessing component, which is responsible for cleaning and formatting the data.

• **Preprocessing**: Preprocessing is performed, as the data received from the devices may contain errors, missing values, duplicate data, or data that requires adjustment, which may compromise or make the ML process unfeasible. Ontologies help in this step. After preprocessing, the AutoML component can now consume the data.

---

[2] Hillar, G.C. **MQTT Essentials-A Lightweight IoT Protocol**. Packt Publishing Ltd,2017.

- **AutoML**: Processes the ML and knowledge discovery algorithms. This component trains and evaluates several ML models and selects the most appropriate model to be used for the given dataset. This model is then used to analyze new data from the equipment, environmental data and other historical data to estimate the probability of certain classes occurring, such as equipment failures. A new predictive model can be selected and adjusted for each new input dataset. As a result, it is possible to maintain solutions generated by the suite of services for each specific DT regarding the predictive model used. After analysis by the Auto ML component, the data is labeled by the knowledge discovery algorithm. The Ontology Model component processes the enriched information.

- **Ontology Model**: The ontology models the entities and their properties for a given context. Based on the relationships between the properties, the Inference Machine component can infer new properties and enrich the dataset once again.

- **Inference Machine**: The Inference Machine component is related to the Ontology Model, as it processes data related to critical events instantiated in ontology, together with the contextual information acquired by the environmental sensors and, based on logical rules, produces new information and relationships that can be used to activate the Autonomous Agent component.

- **Autonomous Agent**: Trigger alerts or modify the System Environment (Figure 1). The Autonomous Agent selects the available devices that should receive the alerts and/or process changes directly in the devices or the operation of the Autonomous Agent, such as turning off a specific machine, for example.

- **Web Interface**: Component that presents information resulting from processing by ML and ontological processing on a dashboard. This provides access to information about the latest critical events detected and displays alert messages that influence predictive maintenance planning. It establishes communication via mobile devices to operators, for example sending an alert for necessary intervention in equipment.

An example of these components' processing will be discussed in Section 4.

### 3.4 Implementation Strategies

The Python programming language was used to specify DT-Create. The PyCaret library was used to create predictive models using ML techniques. The OWL2Ready library was used to manipulate ontologies. Data storage is managed using the MySQL Database Management System (DBMS). The HTML and JavaScript languages were used to develop the Web Interface component. The code is available at [3].

---

[3] https://github.com/izaqueesteves/self_adaptive_article_2024_ufjf

FR01 and FR02 were achieved through the Broker specified in the System Environment component. FR03 was achieved through the Auto ML, Ontology Model, and Inference Machine components. FR04 was achieved through data storage in a database. FR5 was tackled through the Web Interface component. FR6 can be achieved through better decision-making support, thus prioritizing the human factor. In addition, the actions of the Autonomous Agent component help with sustainability because it contributes to a longer lifespan of equipment by preventing critical failures. Functional requirements FR07 and FR08 were met through the evolution of DT-Create, with self-adaptive features and selection of learning models.

Finally, non-functional requirements NFR01 and NFR02 were achieved through the Broker module of the System Environment component. NFR03 were achieved using the AutoML component.

## 4 Evaluation

In DSR, rigor is developed by applying the fundamentals and methods necessary for the research. As a result, scientific knowledge is produced in each of the cycles, serving as a basis for new versions of the solution. This construction of versions is an iterative process. In addition, each iteration refines the artifact to produce better results.

We conduct the study using data from sensors and equipment, which were processed and analyzed to verify how this enriched data supports decision-making regarding predictive maintenance. For this, we adopted the following steps: definition of the scope, detailing of the evaluation scenarios, specification of ontologies, Case Study 1 and Case Study 2 execution and results presentation.

### 4.1 Scope Definition

We defined the scope of the evaluation through GQM (Goal, Questions, Metrics) (Basili and Weiss, 1984). The **objective** is to analyze the support for decision-making **related to** predictive maintenance, **considering** data from sensors and equipment, **in the context** of industry. The Research Question (RQ) to be investigated in this evaluation is: **"How can decision-making related to predictive maintenance be supported in industry using DTs?"**

The evaluation aims to verify whether the artifact works, whether the theoretical conjectures are aligned with expectations, and whether the artifact (DT-Create) assists specifying DTs to support predictive maintenance decision-making. Based on the results of the evaluation in the first cycle and on the scientific knowledge acquired, we compared the theoretical conjectures raised during the literature review. We carried out another iteration, improved the artifact, and re-evaluated. Finally, we verified whether the artifact answered the research question (RQ).

According to the theoretical framework, solutions found in systematic mapping support a complex domain whose context can change quickly. In the industrial context, new sensors can be installed in machines and equipment and new IoT devices can be added. Despite this, the solutions must continue to work properly, supporting maintenance managers in decision-making. The solution must also continue to offer self-adaptation of the production system, even when the application scenario changes and, in turn, the dataset changes. The solution must therefore allow the specification of DTs to support predictive maintenance, despite variations in application scenarios.

In the first evaluation cycle, we specified a DT derived from DT-Create to process data from environmental sensors and failures. Using this DT, we conducted a case study to evaluate the artifact. In the second cycle, we improved the intelligent processing component, and conducted the second case study, specifying a new DT from the improved DT-Create. The case studies refer to scenarios where the solution provided decision support in predictive maintenance, based on intelligent techniques on equipment failure and context data. In both cycles, specific ontologies were created and instantiated. In Case Study 1, the SmartMaintenance ontology was specified. In Case Study 2 the SensorEquipment ontology was specified.

### 4.2 Evaluation Scenarios

The case studies were conducted according to the following steps (Runeson and Martin, 2021) (Runeson et al., 2012): (i) case study design (preparation and planning for data collection), (ii) execution (evidence collection), (iii) analysis of the collected data, and (iv) reporting of the results. The case study scenarios are a textile industry producing fabrics (Case Study 1) and a metallurgical industry supplying heat treatment processes (Case Study 2).

In Case Study 1 (CS1) (DSR Cycle 1), historical data on textile industry machinery failures were used, such as the type of failure, time of the failure event, repair time, cost, failure criticality, temperature, and humidity of the environment at the time of the failure event. The collected data are available on GitHub[3] to facilitate the reproducibility of this case study. In the second cycle, we conducted Case Study 2 (CS2) using machine status data collected from heat treatment furnaces in a large industry. The machines have sensors that collect information about overall machine vibration, collision, imbalance, electrical network, and bearings 1, 2, and 3. During the study, we had direct contact with experts from the metalworking industry. Technical interviews were conducted, considering that the focus was to understand how the data could support decision-making and which critical operational scenarios needed to be mapped[3].

## 4.3 Case Study 1 (First DSR Cycle)

### 4.3.1 Execution

The textile industry is one of the major players in Brazilian industry. Maintaining productivity in the textile industry is down to the automation of machines and industrial processes, among other factors. Using techniques that accelerate decision-making in this sector is strategic to increase productivity. However, its continuity still depends on strategies that ensure their constant innovation, including sensors for data collection and the increasing use of data analysis techniques.

We used DT-Create to specify a DT for the textile production system context. The initial results support the predictive detection of machine failures and the consequent reduction in operating costs. Based on data analysis, DT could trigger alerts. Machines in the textile industry, especially in the mesh sector, present failures during operations that are usually solved quickly, returning the machine to an operational state. However, in the medium term, these failures can indicate operational inefficiency. Therefore, evaluating the criticality of failures through historical analysis of events and environmental data can assist in the predictive maintenance of equipment. In the production context, a tolerated failure may, in the short term, be responsible for a critical failure or even for the machine to stop. For example, a succession of needle breaks in a machine may directly contribute to the failure of a mechanical component and compromise a production batch.

Associated with failure events, there is information from the environment that indirectly contributes to the occurrence of critical failures. For example, the temperature and humidity of the production environment interfere with the yarns' properties and affect the machines' operating conditions. In addition, the proximity and experience of a given operator may be crucial in detecting the probability of failure or in analyzing the environment for the occurrence of a failure.

To conduct this Case Study, we used a dataset with 10,000 failure records and other technical and managerial information. The dataset has variables such as machine identification, type of failure, timestamp, repair team, repair cost, humidity and temperature. Regarding the data for supervision, there is the class in which the failure was classified according to the industry's own perception: class 1 for critical failures and 0 for non-critical failures. Failure records for training also have criticality signaling, which favors the supervised learning strategy. Figure 2 shows the input dataset.

|  | machine_id | type_of_failure | timestamp | time_repair | cost | criticality | humid | temp | label |
|---|---|---|---|---|---|---|---|---|---|
| 0 | 81 | 7 | 1583193600 | 0.363742 | 0.522 | 0.499 | 61 | 97 | 0 |
| 1 | 66 | 5 | 1593388800 | 0.297000 | 0.485 | 0.307 | 63 | 112 | 0 |
| 2 | 32 | 9 | 1600128000 | 0.383106 | -0.075 | 0.798 | 6 | 40 | 1 |
| 3 | 43 | 4 | 1597708800 | 0.300732 | 0.553 | 0.531 | 48 | 141 | 0 |
| 4 | 62 | 2 | 1585699200 | 0.127000 | 0.335 | 0.367 | 74 | 148 | 0 |
| ... | ... | ... | ... | ... | ... | ... | ... | ... | ... |
| 9994 | 34 | 3 | 1599350400 | 0.391921 | 0.518 | 0.508 | 6 | 56 | 0 |
| 9995 | 78 | 3 | 1598918400 | 0.102000 | 0.494 | 0.272 | 56 | 88 | 0 |
| 9996 | 81 | 9 | 1595376000 | 0.339469 | 0.277 | 0.456 | 80 | 142 | 0 |
| 9997 | 97 | 7 | 1600905600 | 0.277222 | 0.457 | 0.424 | 61 | 99 | 0 |
| 9998 | 44 | 9 | 1581206400 | 0.250000 | 0.309 | 0.282 | 42 | 92 | 0 |

Figure 2 - Dataset with failure data.

A failure criticality classification model was implemented. The goal was to automatically classify failures as critical or not. The dataset with 10,000 failure records was organized into two sets: one with 75% of the data for model training and the remaining 25% for evaluating the model's accuracy in class predictions. The dataset was preprocessed to ensure data normalization.

ML techniques were performed using a neural network based on the Perceptron multilayer classifier (Rosenblatt, 1962), which has one or more hidden layers with a certain number of neurons and is trained using a backpropagation algorithm. In this way, the neural network can automatically classify new failure records, which are instantiated in the ontological model. To evaluate the model accuracy, we performed cross-validation, which resulted in an accuracy of 99%, as shown in Figure 3.

```python
from sklearn.model_selection import cross_val_score

# Perform cross validation
scores = cross_val_score(clf, y_test, y_aux, cv=8)  # cv is the number of folds

# View the results
print("Accuracy scores for each fold:", scores)
print("Mean cross-validation score:", scores.mean())
```

```
Accuracy scores for each fold: [0.99680511 1.         0.99679487 0.99679487 1.         0.99679487
 0.99679487 0.99679487]
Mean cross-validation score: 0.9975974338494307
```

Figure 3 – Predicted Model cross-validation.

Additionally, we evaluated the performance of the classification model considering how the classes of interest are determined. After testing, a confusion matrix was generated (Figure 4). The occurrence of false positives and false negatives can compromise the accuracy, and the confusion matrix shows the distribution of predictions within the four categories: true positives, false positives, true negatives, and false negatives. By analyzing the class distribution between predicted and accurate labels, we observed that the model has high accuracy in predicting both positive and negative classes. The high accuracy can be explained by factors such as the size of the dataset available for testing the model and the imbalance between the positive and negative classes.

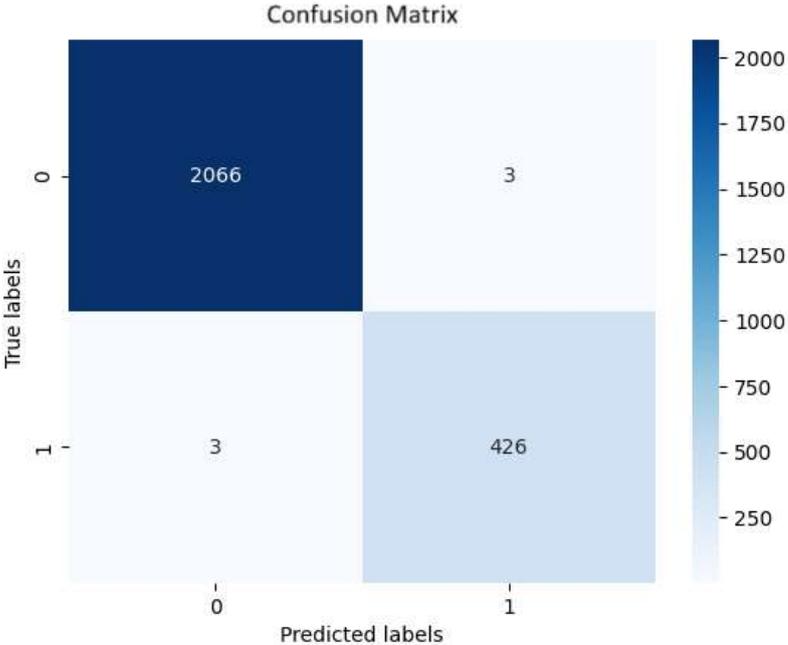

Figure 4 – Confusion Matrix for the predictive model.

The result of processing via a neural network was a new dataset with defined classes. The resulting dataset with the assigned criticality values was processed by the SmartMaintenance ontology (detailed in the following subsection). The DT used in Case Study 1 favors this by providing data from ML, combined with context data, and processed through ontologies and autonomous software agents. Thus, this scenario considers three important elements for failure management and maintenance: building knowledge about failure, monitoring imminent critical failures, and self-control of production equipment through autonomous software agents.

### 4.3.2 SmartMaintenance Ontology

The ontology was developed to support the semantic processing of information in the predictive maintenance of failures in textile equipment (purpose). The scope was defined considering the predictive maintenance domain (scope). The OWL 2.0[4] and SWRL[5] languages were used as implementation languages, and the OWLReady2 library was used to import data from a relational database into the ontology. The users are decision-makers in the industry domain of predictive maintenance.

---
[4] https://www.w3.org/TR/owl2-overview/
[5] https://www.w3.org/submissions/SWRL/

With the help of experts, the Competency Questions (CQ)[6] were defined as follows: (i) What is the average temperature at the time of failure occurrence? (ii) Do failures occur more frequently in high humidity conditions compared to low humidity conditions? (iii) What type of failure is more common in certain environmental conditions? (iv) Are there differences in failure between work shifts? (v) What is the average criticality of failures in high-temperature versus low-temperature conditions?

Logical rules related to the ontology terms were specified to answer the QCs. The main classes and relationships (*objectProperties*) of the ontology are presented in Figure 5. The *Failure* class represents the failure event containing information such as the equipment failure code, the sector to which the equipment belongs, and the code that identifies the failure event. *FailureType* is a subclass of *Failure*, representing the type of failure. Associated with Failure class is the *DateTime* class that provides the timestamp of the failure event. The failures occur in a production environment where the humidity and temperature variables are controlled, which are properties of the Failure class.

Criticality class represents the degree of criticality estimated by the intelligent processing component of a failure. The criticality of a failure can be projected based on historical data, costs, repair time, and others. The *Alert* class is associated with the *Failure* class through the *hasAlert* objectProperty, which is dynamically defined based on Semantic Web Rule Language (SWRL) rules that consider specific characteristics of failure occurrences, such as the number of a given event classified as critical, and environmental data such as temperature and humidity. Figure 5 shows also the main objectProperties in the Protegé[7].

When logical rules are established for ontological processing, real-time context information enriches information and leverages the knowledge processed by ML algorithms. This combination is distinctive in managing maintenance processes and provides decision support for predictive maintenance interventions. Table 1 shows the main SWRL rules, according to operational parameters defined by experts, which consider the environment's temperature and humidity indices.

---

[6] The use of competence questions (CQs) is used as a means to define the ontology requirements and help identify the necessary concepts, properties, and relations. CQs are questions that the ontology should be able to answer. Thus, they provide a mechanism to verify if the ontology is in accordance with the established requirements and properly represents the desired knowledge.

[7] https://protege.stanford.edu/

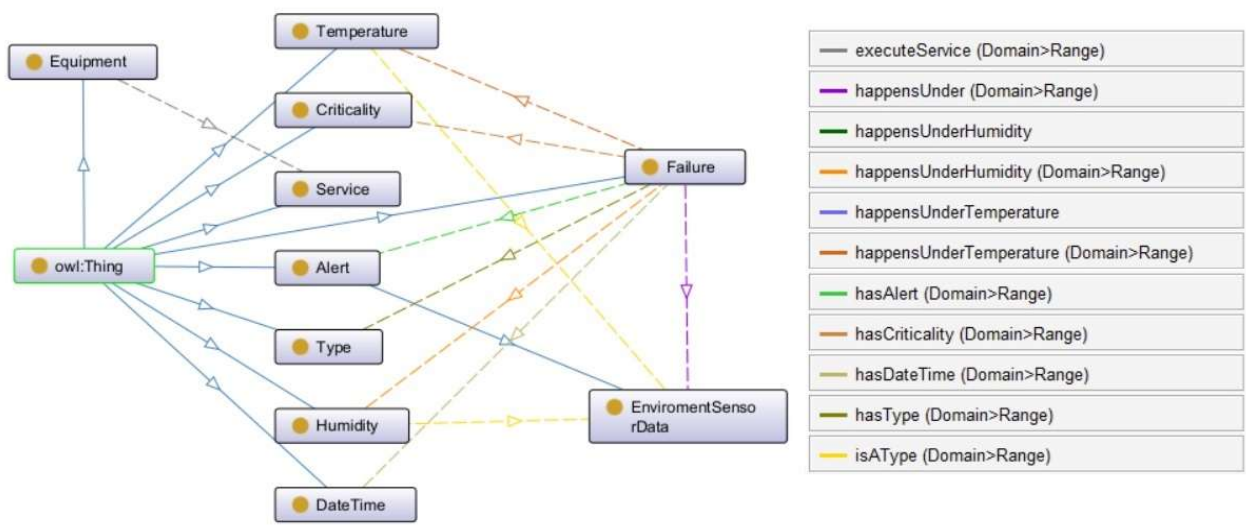

Figure 5 - Classes and properties of the SmartMaintenance ontology.

Table 1 – SWRL rules for evaluating temperature and humidity

| # | SWRL rules | Description |
|---|---|---|
| 01 | **Antecedent**<br>smartmaintenance:Failure(?f) ^<br>smartmaintenance:Alert(?a) ^<br>smartmaintenance:Humidity(?h) ^<br>smartmaintenance:humidityValue(?h, ?humid) ^<br>smartmaintenance:Temperature(?t) ^<br>smartmaintenance:temperatureValue(?t, ?temp)<br>^ swrlb:lessThanOrEqual(?humid, 25) ^<br>swrlb:greaterThanOrEqual(?temp, 35)<br><br>**Consequent**<br>-> smartmaintenance:alertCode(?f, 100) | If a failure occurs and a condition is detected where the humidity is less than or equal to 25%, and the temperature is greater than or equal to 35°C, then associate alert code 100 with that failure. |
| 02 | **Antecedent**<br>smartmaintenance:temperatureValue(?t, ?temp) ^ smartmaintenance:Failure(?f) ^<br>swrlb:greaterThanOrEqual(?temp, 30) ^<br>swrlb:equal(?type, 4) ^<br>smartmaintenance:Temperature(?t) ^<br>smartmaintenance:numberOfOccurrences(?f, ?num) ^ smartmaintenance:typeOfFailure(?f, ?type) ^ swrlb:greaterThanOrEqual(?num, 4)<br>**Consequent**<br>-> smartmaintenance:alertCode(?f, 200) | If there is a temperature (*Temperature*) associated with a value (*temperatureValue*) that is greater than or equal to 30, and there is a failure (*Failure*) with a specific type (*typeOfFailure*) equal to 4, and this failure has occurred at least 4 times (*numberOfOccurrences* greater than or equal to 4), then assign alert code 200 (*alertCode*) to this failure. |

Considering the dataset, the ontology processed the combination of the failure data already classified by the neural network with the environmental data and qualitative and quantitative failure metrics that should be observed. Figure 6 details an example of how the ontology processed the inference of properties of the "Failure1" instance with accumulated failures and failure type attributes set respectively as "4" and "5".

Considering the environmental information (temperature and humidity), the SWRL rules also evaluated the scenarios where a technical intervention was urgent to mitigate the risk. Rule 01 in Table 1 generated an alert code with a value of 100 when the ambient temperature is higher than 35°C, and the ambient humidity is lower than 25%. In another example of using a SWRL rule, now processed by rule 02 in Table 1, there is an instance "Fault1", defined with the following attributes and values: temperature: 45°C, humidity: 20%, failure type: 4, and number of failures occurrences: 5. After running the inference, the ontology processed two alerts with codes "100" and "200", as shown in Figure 7.

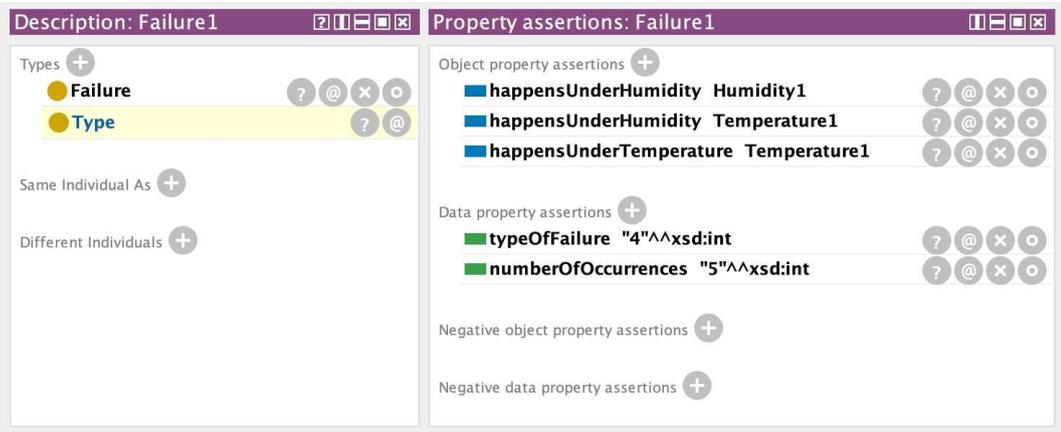

Figure 6 - Properties inferred by the ontology (Protegé software)

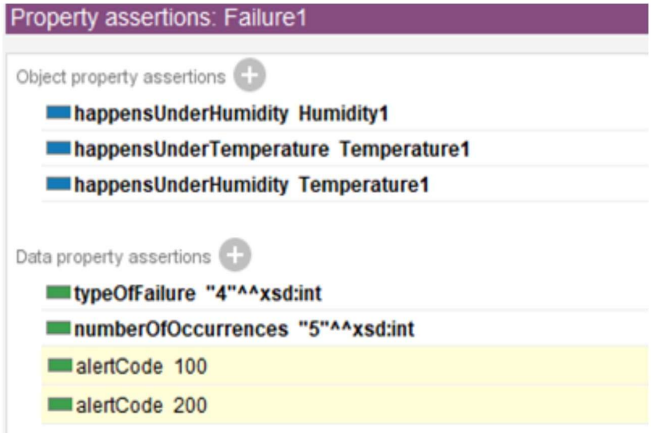

Figure 7 - SWRL rule for alert generation represented in the Protegé software.

### 4.3.3 Analysis

From the textile machine data and environmental data, processed by a neural network and subsequently instantiated in the SmartMaintenance ontology and processed by logical rules, it was possible to obtain insights related to predictive maintenance. For example, most critical failures were directly related to environmental humidity, i.e., the alerts generated by the ontology involving high humidity conditions affected failure events classified as critical by the predictive model. The inferences processed indicate the need to send alerts from the autonomous agent to nearby operators to check specific items on the machine, such as needles that could break, in a particular case. The solution was to interrupt the operation before more critical failures could compromise the operation of the equipment.

Therefore, the initial results support predictive detection of machine failures and consequent reduction in operating costs. Based on the data analysis, the DT created triggered alerts, providing an intelligent and contextual solution for Industry 5.0 where the human factor and sustainability are highlighted.

However, we detected that messages can be sent directly to the machines or IoT devices responsible for the machines through MQTT, according to the alert generated, to preserve the equipment or even ensure safe conditions for operation. Another critical point is the exclusive use of a neural network algorithm. Moreover, depending on the type of data coming from sensors and the variables used, a neural network may not be the best ML technique, given each dataset's characteristics and performance issues. Therefore, for the next development cycle (DSR cycle) of the artifact, the possibility of choosing the most appropriate ML technique to be used can be proposed.

#### 4.3.4 Results

The solution proposed in the first DSR cycle was able to jointly evaluate the criticality of failures associated with environmental data and the qualitative and quantitative characteristics of failures, enriching the information. It also produced data that allowed the autonomous agent to act by sending alert messages to machine operators. Therefore, the combined use of semantic processing and machine learning allowed autonomous software agents to detect maintenance-related failures. However, we realized that it was necessary to evolve the artifact concerning the selection of a predictive model for data processing and self-adaptation. Then, based on the identification of this point of improvement, a new DSR cycle was conducted.

### 4.4 Case Study 2 (Second DSR Cycle)

#### 4.4.1 Execution

The metallurgical industry uses heat treatment furnaces in its mechanical manufacturing processes. These are industrial equipment capable of altering the physical and, in some cases, chemical properties of materials through the controlled application of heat. These furnaces operate over a wide range of temperatures. They are essential in tempering, annealing, hardening, and normalization among others, and are applied mainly to metals and alloys to improve their resistance, flexibility, toughness, and other characteristics essential for the specific application of the material.

Considering condition-based maintenance (CBM), vibration analysis is one of the techniques used. This is based on monitoring and analyzing the vibrations emitted by equipment in operation. Each machine has a vibration pattern that is considered "normal" when it is operating properly. Deviations from this pattern may indicate the presence of failures or wear in components, such as bearings, gears, shafts, and even structural problems in the furnace itself or its support and movement systems.

Recently, the predictive maintenance area of a metalworking plant underwent digitalization of processes and now uses sensors in a large part of the production assets. In other words, a large volume of data is collected daily, and there is an opportunity for more efficient planning of maintenance interventions. Although there are continuous monitoring systems for machines, anticipating failures, i.e., predictive maintenance considering the condition of the machines, is one of the objectives. Therefore, better support for furnace operators is possible, based on more informative dashboards that support decision-making more easily, as well as the possibility of automatically shutting down equipment in a preventive manner. Another highlight is the industrial sustainability, since furnaces are important equipment in production. Shutdowns for unscheduled repairs consume a considerable number of resources and need specialized personnel, in addition to compromising the production schedule.

A dataset with data from sensors related to specific furnaces was provided by a known international industry. One of its divisions is in the state of Paraná, Brazil. For this study, the metallurgical industry provided data with 1,000,000 records from three different machines containing information from seven sensors, which indicate specific failure modes, installed in heat treatment furnaces for metal parts, including global vibration, collision, imbalance, network, bearing 1, bearing 2, bearing 3. The dataset represents an operational interval of 8 days. The set of sensors is used to monitor the vibration of the equipment during its use. For each sensor, the company establishes value ranges, which represent intervals considered acceptable or normal, based on the knowledge of its experts, as shown in Table 2.

Table 2 – Intervals considered normal for data captured by sensors.

| Sensor | Lowest Value | Highest Value |
|---|---|---|
| Overall vibration (S1) | 0 | 2.5 |
| Collision (S2) | 0 | 5000 |
| Unbalance (S3) | 0 | 2.5 |
| Network (S4) | 0 | 5 |
| Bearing 1 (S5) | 0 | 50 |
| Bearing 2 (S6) | 0 | 60 |
| Bearing 3 (S7) | 0 | 360 |

The data did not present class information that would allow the learning algorithm to predict which class a piece of data is associated with. With improved DT-Create, other machine-learning algorithms can be considered and selected. In the specific case of the metallurgical industry data, unsupervised algorithms were used to detect patterns in the data, and, considering the characteristics of the dataset provided, it was not possible to extract information that would favor decision-making in predictive maintenance activities. Therefore, before starting the machine learning tasks, a procedure was executed to estimate the machine failure condition from the sensor failure conditions, a special pre-processing. With this the available data could be classified, and then a predictive model could be trained to classify new data.

From the available data and using statistical procedures, it was possible to estimate whether the value observed by the sensor indicated the occurrence of a failure mode (Bussab and Morettin, 2010). To do so, we followed these steps for data labeling:

1. Calculate the Confidence Interval (CI) of each observation of each sensor;

2. Estimate the failure condition of each sensor by comparing the calculated CI with the reference values in Table 2;

3. Determine the expected value of the output of each sensor tuple;

4. Compare the expected value with the maintenance management limit.

We considered the sample size and, disregarding the effect of outliers, assumed the normality of the data (Fischer, 2011), which does not generate a loss of generalization. According to Moore (1996), normal distributions are good approximations of the results of many types of random results applied to a wide range of phenomena in the real world, including error measurements in manufacturing processes. Also, many statistical inference procedures based on normal distributions work well for other approximately symmetric distributions. Therefore, we adopted the confidence interval analysis to estimate, with a 95% confidence level, the classification of each output of each sensor as a failure condition (1) or normal condition (0), calculated as follows:

$$IC = x_i \pm z \frac{s}{\sqrt{n}} \qquad (1)$$

Where:
- xi is the sample mean;
- z is the critical value of the standard normal distribution corresponding to the desired confidence level (e.g., 1.96 for a 95% CI);
- s is the sample standard deviation;
- n is the sample size.

Once the confidence intervals for each observation of each sensor were known, it was possible to verify whether the estimated confidence interval was contained within the interval defined in Table 2, producing the label 0 indicating normal condition or label 1, indicating failure condition. Table 3 shows a set of labeled sensors generated by a given observation. Highlighted are the observations outside the normal operating range (sensors S4 and S7).

Table 3 – Labels defined after checking confidence intervals and parameters

| Sensor | Observed Value | Label |
|---|---|---|
| Overall vibration (S1) | 0.093809 | 0 |
| Collision (S2) | 8.133628 | 0 |
| Unbalance (S3) | 0.050851 | 0 |
| **Network (S4)** | **0.069705** | **1** |
| Bearing 1 (S5) | 4.365503 | 0 |
| Bearing 2 (S6) | 5.017564 | 0 |
| **Bearing 3 (S7)** | **49.76829** | **1** |

After this step, we have the set of observed values and the respective set of labels. Based on the operational knowledge about the relevance of each failure mode (sensors) on the machine failure condition, weights were assigned to each of the sensors as shown in Table 4, satisfying:

$$\sum_{Pi} = 1; Pi \geq 0 \qquad (2)$$

where Pi is the sensor's relevance to defining the machine failure condition, with estimated values in the interval [0,1].

Table 4 – Relevance of each sensor in determining the condition of a machine.

| Sensor | Weight |
|---|---|
| Overall vibration (S1) | 0.10 |
| Collision (S2) | 0.03 |
| Unbalance (S3) | 0.10 |
| Network (S4) | 0.02 |
| Bearing 1 (S5) | 0.25 |
| Bearing 2 (S6) | 0.30 |
| Bearing 3 (S7) | 0.20 |

Combining the tuple of labels of each sensor with the respective weights, it was possible to calculate the expected value for the machine failure condition, exemplified in Table 5. The expected value is obtained by:

$$E = \sum_{Si.Pi} \qquad (3)$$

where Si represents the label (0 or 1) associated with a sensor, and Pi is the weight associated with the sensor.

Table 5 – Combination between label weight to estimate relevance.

| Sensor | Weight | Label | Weight x Label |
|---|---|---|---|
| Overall vibration (S1) | 0.10 | 0 | 0 |
| Collision (S2) | 0.03 | 0 | 0 |
| Unbalance (S3) | 0.10 | 0 | 0 |
| Network (S4) | 0.02 | 1 | 0.02 |
| Bearing 1 (S5) | 0.25 | 0 | 0 |
| Bearing 2 (S6) | 0.30 | 0 | 0 |
| Bearing 3 (S7) | 0.20 | 1 | 0.20 |

The sample presented in Table 5 highlights an expected value equal to 0.22. The expected value can be used as a reference for decision-making on maintenance interventions according to the machine or the production demand based on scenarios. For example, considering a machine with a long maintenance history, i.e., a machine more sensitive to failures, the expected value can be adjusted close to zero, i.e., in the occurrence of expected values close to zero, the failure condition is already signaled. On the other hand, a recently acquired machine with a low maintenance rate can be operated under less conservative conditions, i.e., the warning signaling considers failure modes with a higher weighted value. Table 6 shows three examples of scenarios for three different machines.

Table 6 – Expected value adjustment scenarios.

| Maintenance Condition | Management Style | Expected Value (EV) |
|---|---|---|
| High | Conservative | 0 < EV < 0.3 |
| Medium | Moderate | 0.3 <= EV < 0.6 |
| Low | Aggressive | EV >= 0.6 |

Therefore, after executing these data pre-processing tasks, the sensor dataset was enriched, which allows us to infer the machine failure condition, as shown in the example in Table 7. If we consider the management parameters in Table 6 and, based on the data in Table 7, the machine failure condition indicates the need for intervention in a conservative scenario.

Table 7 – Machine condition estimation from sensor data.

| S1 | S2 | S3 | S4 | S5 | S6 | S7 | Failure Condition |
|---|---|---|---|---|---|---|---|
| 0.09 | 8.13 | 0.05 | 0.07 | 4.37 | 5.01 | 49.7 | 0,2 |

The sensor data were enriched with labels 0 or 1, indicating, according to the management style parameters shown in Table 7, the need for intervention for a given machine. Figure 8 shows what the dataset is like after statistical preprocessing, with a new column added, named "label", indicating the need for intervention (1) or not (0).

|  | 2F03_COLISÃO_S3 | 2F03_VIBRAÇÃO_GLOBAL_S3 | 2F03_DESBALANCEAMENTO_S3 | label |
|---|---|---|---|---|
| 63023 | 15.892150 | 0.245994 | 0.065820 | 1 |
| 63271 | 6.978260 | 0.239681 | 0.065876 | 1 |
| 248179 | 11.011000 | 0.352170 | 0.057822 | 1 |
| 248180 | 10.790450 | 0.352170 | 0.057822 | 1 |
| 248181 | 15.189050 | 0.352170 | 0.057822 | 1 |
| ... | ... | ... | ... | ... |
| 239509 | 11.859380 | 0.150089 | 0.061494 | 0 |
| 239510 | 10.462910 | 0.150089 | 0.061494 | 0 |
| 239511 | 17.304460 | 0.150089 | 0.061494 | 0 |
| 239512 | 13.316650 | 0.150089 | 0.061494 | 0 |
| 313731 | 9.948064 | 0.136900 | 0.059801 | 0 |

Figure 8 - Adding a column to the dataset after statistical processing.

With the dataset properly labeled and before ML processing, it was possible to see that the dataset presents a large imbalance between classes, that is, there is a large disparity between data with label 0 and data with label 1. Figure 9 shows that in the sample time interval, class 0 is predominant over class 1 (abnormality).

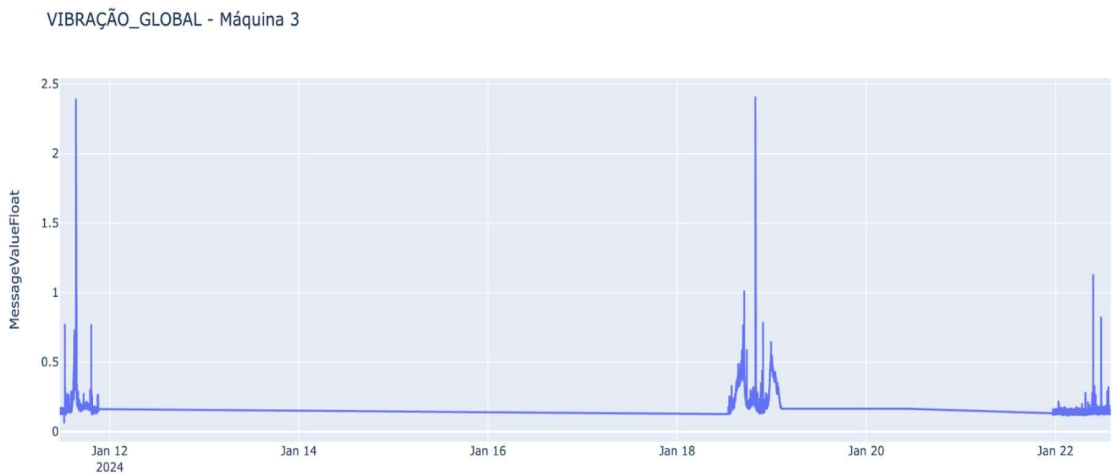

Figure 9 - Sample data from a sensor.

Class imbalance is a common issue in ML classification problems. It occurs when the target classes (the categories the model tries to predict) are unevenly represented in the data. This means that one or more classes have a much larger number of instances (examples) than others in the dataset. Class imbalance can lead to several problems during the training and evaluation of machine learning models, negatively influencing the model's performance, especially in its ability to predict the least represented classes. In this specific case, the imbalanced class is precisely one of the most relevant (He and Garcia, 2009). Because of this, we adopted under-sampling of the majority class, which reduces the number of instances in the majority class to balance the distribution of the classes (Chawla et al., 2002).

With the pre-processed dataset, it was possible to submit it to the DT-Create suite of services to select the learning model and then specify the DT used in this Second Case study. This automatic selection chose one of the available machine learning models based on the performance of each on the dataset. This automatic selection is performed in DT-Create using the PyCaret library[8].

```
from pycaret.classification import * test=data.sample(frac=0.10, random_state=1)
train=data.drop(test.index)
```

```
reg = setup(data = train, target = 'label', remove_outliers= True
, normalize = True, feature_selection = True, session_id=1) best = compare_models()
```

The process began by dividing the dataset into training and test data. Then, the model creation was parameterized. During parameterization, the dataset was treated with outlier removal, data normalization, and selection of the most relevant features. The model comparison evaluates the performance in both predictive accuracy and computational cost, and as output, an ordered list of classification models is generated. The results in shown in Figure 10.

---

[8] PyCaret is a machine learning library in Python that automates the ML workflow. At the end of the machine learning process, the expected output is a model adjusted for prediction based on the input of new machine condition data, as presented in the code excerpt below.

|   | Model | Accuracy | AUC | Recall | Prec. | F1 | Kappa | MCC | TT (Sec) |
|---|---|---|---|---|---|---|---|---|---|
| dt | Decision Tree Classifier | 0.9526 | 0.9898 | 0.9514 | 0.9813 | 0.9659 | 0.8883 | 0.8902 | 0.6240 |
| et | Extra Trees Classifier | 0.9526 | 0.9898 | 0.9514 | 0.9813 | 0.9659 | 0.8883 | 0.8902 | 0.7740 |
| lightgbm | Light Gradient Boosting Machine | 0.9480 | 0.9919 | 0.9460 | 0.9802 | 0.9625 | 0.8782 | 0.8808 | 0.8290 |
| rf | Random Forest Classifier | 0.9473 | 0.9908 | 0.9481 | 0.9770 | 0.9620 | 0.8757 | 0.8781 | 0.8340 |
| knn | K Neighbors Classifier | 0.9427 | 0.9854 | 0.9481 | 0.9707 | 0.9588 | 0.8646 | 0.8672 | 0.4450 |
| gbc | Gradient Boosting Classifier | 0.9404 | 0.9894 | 0.9416 | 0.9734 | 0.9570 | 0.8598 | 0.8620 | 0.5220 |
| xgboost | Extreme Gradient Boosting | 0.9373 | 0.9883 | 0.9286 | 0.9821 | 0.9542 | 0.8551 | 0.8600 | 0.4570 |
| ada | Ada Boost Classifier | 0.9045 | 0.9624 | 0.9112 | 0.9522 | 0.9309 | 0.7763 | 0.7796 | 0.7010 |
| nb | Naive Bayes | 0.6792 | 0.6946 | 0.7489 | 0.7882 | 0.7673 | 0.2492 | 0.2515 | 0.4160 |
| qda | Quadratic Discriminant Analysis | 0.6792 | 0.6946 | 0.7489 | 0.7882 | 0.7673 | 0.2492 | 0.2515 | 0.4100 |
| svm | SVM - Linear Kernel | 0.6646 | 0.0000 | 0.7620 | 0.7737 | 0.7532 | 0.1856 | 0.2052 | 0.4060 |
| ridge | Ridge Classifier | 0.6158 | 0.0000 | 0.6504 | 0.7710 | 0.7048 | 0.1634 | 0.1695 | 0.4060 |
| lda | Linear Discriminant Analysis | 0.6158 | 0.6558 | 0.6504 | 0.7710 | 0.7048 | 0.1634 | 0.1695 | 0.4050 |
| lr | Logistic Regression | 0.6127 | 0.6558 | 0.6461 | 0.7699 | 0.7018 | 0.1590 | 0.1653 | 0.5910 |
| dummy | Dummy Classifier | 0.2941 | 0.5000 | 0.0000 | 0.0000 | 0.0000 | 0.0000 | 0.0000 | 0.4100 |

Figure 10 - Comparison between machine learning models.

After comparing the models, the most suitable model was the Decision Tree. Despite being close to the other models, it stands out for its accuracy, recall, F1Score, Kapp, and MCC metrics[9].

Once the most suitable classification model was found, the next step was to define it and adjust the hyperparameters. The fold parameter specifies the number of subsets (folds) into which the data set will be divided during cross-validation.

```
dt = create_model('dt',fold=3) tuned_dt =
tune_model(dt,optimize="Accuracy") tuned_model_custom =
tune_model(dt)
 predict_model(dt)
```

---

[9] Accuracy evaluates the overall performance of a classification model by showing the proportion of correct predictions (both positive and negative) about the total predictions made. Recall, also known as sensitivity or true positive rate, evaluates the performance of a classification model, especially in contexts where classes are imbalanced or when the detection of minority classes is particularly important. It measures the proportion of real positive instances that the model correctly identified, that is, the fraction of true positives about the total number of effectively positive instances (the sum of true positives and false negatives). The F1-Score evaluates the performance of models, especially in situations where there are imbalanced classes or when precision and recall are equally important. It combines precision and recall into a single metric through their harmonic mean, balancing these two measures. The Kappa coefficient is a metric used to measure the agreement between two evaluations, quantifying how much the classification made by a machine learning model agrees with the true classification (or to another assessment ) beyond what would be expected by chance. The Matthews Correlation Coefficient (MCC) evaluates the quality of binary classification models. It can be used even when the classes are of very different sizes, making it particularly useful in imbalanced data sets.

After creation, the model was adjusted using the tune_model function (code excerpt above). Here, the adjustment is made to increase its generalization capacity. Then, the adjusted model (dt) was defined as the prediction model (predict_model). We have the adjusted prediction model as output, as shown in Figure 11. The adjusted model has a lower performance than the model initially chosen. This occurs because the adjustment reduces the overfitting of the model, which means that the model performs well with training data but could perform better when predicting results on unseen or test data. Therefore, even though it is less accurate than the initial model, it is more precise in predicting new data after adjustment.

|   | Model | Accuracy | AUC | Recall | Prec. | F1 | Kappa | MCC |
|---|---|---|---|---|---|---|---|---|
| 0 | Decision Tree Classifier | 0.9199 | 0.9776 | 0.9093 | 0.9757 | 0.9413 | 0.8158 | 0.8208 |

Figure 11 - Fitted machine learning model.

After choosing the most appropriate machine learning model from DT-Create, the model was used to specify the DT to support decision-making and predictive maintenance procedures. We created an ontology specifically for the Case Study 2 application domain based on Competence Questions (CQ) for semantic processing.

### 4.4.2 SensorEquipment Ontology

The SensorEquipment ontology is aimed at predictive maintenance of machines based on the analysis of operating conditions indicated by sensors. For its specification, Competency Questions were elaborated on with the help of experts: (i) Which machines currently indicate an error condition? (ii) Which sensors indicated error conditions on a specific machine? (iii) Are there error patterns that co-occur on several machines? (iv) What were the last readings of each sensor on a specific machine before a maintenance event? (v) How many maintenance events were initiated due to error conditions in the last six months? (vi) Which machines did not indicate an error in the last 12 months? (vii) What is the average frequency of errors for each type of sensor on different machines?

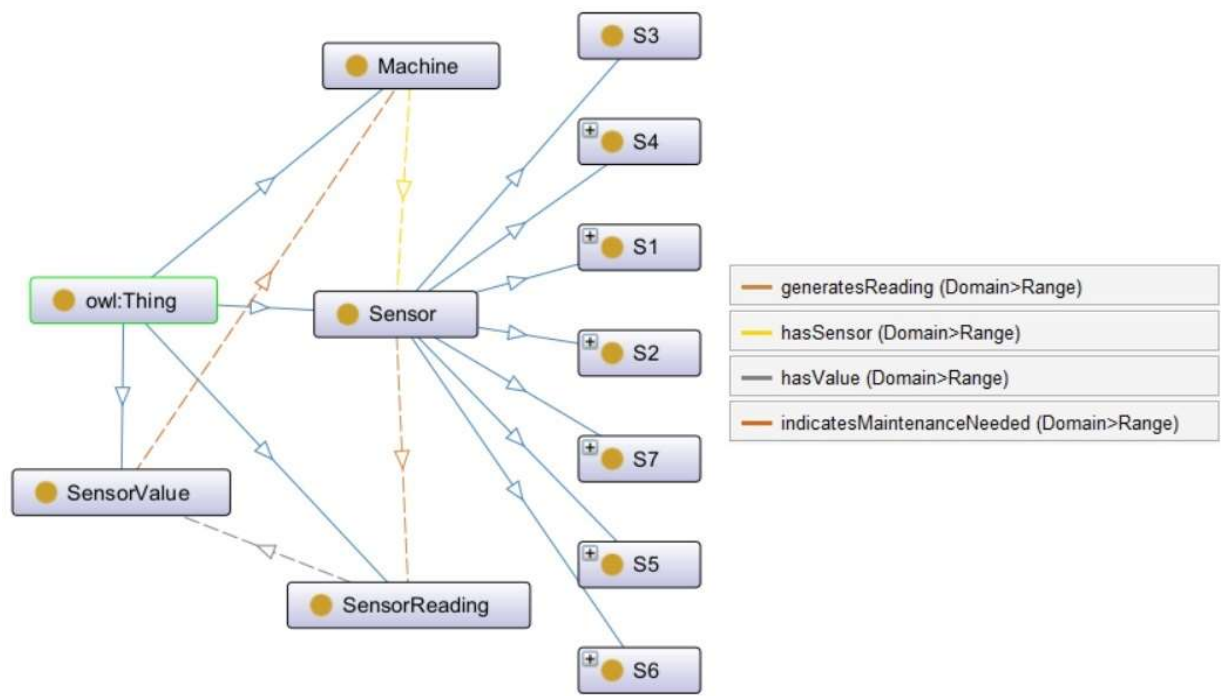

Figure 12 - Ontology for representing the operational condition of machines.

Figure 12 presents the main classes of the ontology, detailed below:

- **Machine:** Represents the machines (generically).
- **Sensor:** Represents the machine's sensors.
- **SensorReading:** Represents the acquisition of data from a sensor.
- **SensorValue:** Values returned by the sensors.

Based on the information on the classes that make up the ontology, the objectProperties that show the relationships between the classes indicated in Figure 12 were described:

- **hasSensor (Machine -> Sensor):** Indicates if a machine has one or more sensors.
- **generatesReading (Sensor -> SensorReading):** Relates a sensor to its readings.
- **hasValue (SensorReading -> xsd:int):** The numeric value of a sensor reading, which can be 0 (normal) or 1 (error).
- **indicatesMaintenanceNeeded (Machine -> MaintenanceEvent):** Indicates that a machine requires maintenance.

The data pre-processed by the ML model were instantiated in the ontology, and the processing of SWRL rules allowed the enrichment of the information. The following SWRL rules were specified to discover insights regarding the opportunities for anticipating maintenance actions from sensor data. As an illustration, to infer the need for maintenance from sensor data, the rules that analyze the readings of the sensors labeled as 0 (value measured within the normal operating range) or 1 (outside the normal operating range) were processed, as shown in Table 7.

Table 7 – SWRL rules for evaluating the machines' operational condition.

| # | SWRL rules | Description |
|---|---|---|
| 01 | **Antecedent**<br>generatesReading(?s1, ?reading) ^ hasSensor(?machine, ?s1) ^ hasValue(?leitura, ?readValue) ^ SensorValue(?readValue) ^ sensorReadingValue(?readValue, 1) ^ Machine(?machine) ^ S1(?s1) ^ SensorReading(?reading)<br><br>**Consequent**<br> -> indicatesMaintenance(?machine, 1) | If a sensor generates a reading whose value is 1 (indicating an error), then the machine associated with that sensor requires maintenance. |
| 02 | **Antecedent**<br>Machine(?machine) ^ S5(?s5) ^ S6(?s6) ^ S7(?s7) ^ hasSensor(?machine, ?s5) ^ hasSensor(?machine, ?s6) ^ hasSensor(?machine, ?s7) ^ generatesReading(?s5, ?reading5) ^ generatesReading(?s6, ?reading6) ^ generatesReading(?s7, ?reading7) ^ hasValue(?reading5, ?readValue5) ^ hasValue(?reading6, ?readValue6) ^ hasValue(?reading7, ?readValue7) ^ sensorReadingValue(?readValue6, 1) ^ sensorReadingValue(?readValue7, 1) ^ sensorReadingValue(?readValue5, 1)<br><br> **Consequent**<br>   -> indicatesMaintenance(?machine, 1) | If the set of bearings that equip a machine, monitored by sensors S5, S6, and S7, show a failure signal simultaneously, an alert is generated. |

Rule 01 of Table 7 considers the occurrence of a failure associated with sensor S1 (global vibration). In this case, sensor S1 is important in the functional context of the machine and needs to be monitored. Rule 02 evaluates the occurrence of failures in a group of bearings that equip a machine, represented by sensors S5, S6, and S7. Specifically, the three simultaneous failure modes signal the need for evaluation by the maintenance team. Thus, from a SPARQL query (code excerpt below), it was possible to retrieve the number of failure events indicated by a group of sensors.

```
SELECT ?sensor (COUNT(?event) AS ?numberOfFailures)
WHERE {
?sensor ex:ErrorInterval ?event .
FILTER (?sensor IN (S5, S6, S7))
}
GROUP BY ?sensor
```

As a result, the ontological processing in Case Study 2 allowed knowledge to be extracted from semantic processing and from verifying the joint occurrence of events important from the point of view of predictive maintenance. This information was not part of the original dataset and was also not determined during processing in the AutoML component.

### 4.4.3 Analysis

With the model adjusted and ready for use, its performance can be verified. Figure 13 shows the predictive model being processed considering new data from sensors.

| | 2F03_COLISÃO_S3 | 2F03_VIBRAÇÃO_GLOBAL_S3 | | 2F03_DESBALANCEAMENTO_S3 | label | prediction_label | prediction_score |
|---|---|---|---|---|---|---|---|
| 726 | 8.522325 | 0.241362 | ... | 0.066188 | 1 | 1 | 0.7298 |
| 348 | 7.289566 | 0.129020 | | 0.060549 | 0 | 0 | 0.6406 |
| 102 | 11.989620 | 0.135731 | | 0.058455 | 1 | 1 | 0.9944 |
| 1272 | 12.550910 | 0.133094 | | 0.061983 | 1 | 1 | 0.9985 |
| 1994 | 17.922621 | 0.388702 | | 0.060862 | 1 | 1 | 0.9979 |
| ... | ... | ... | | ... | ... | ... | ... |
| 1104 | 13.603150 | 0.146172 | | 0.061615 | 0 | 0 | 0.9555 |
| 1179 | 11.143620 | 0.135574 | | 0.062383 | 1 | 1 | 0.9983 |
| 1654 | 9.865036 | 0.136189 | | 0.062684 | 1 | 1 | 0.9722 |
| 368 | 10.959950 | 0.142156 | | 0.060549 | 0 | 0 | 0.6406 |
| 859 | 21.858290 | 0.253324 | | 0.065797 | 1 | 1 | 0.9901 |

Figure 13 - Processing the predictive model with a new data set.

As seen in Figure 11, the model performs well in all evaluated metrics, as indicated by high scores in AUC, Kappa, and MCC. However, the dataset used to create the model is relatively small in relation to the sample period. The input data is expected to be balanced in signaling machine failure and non-failure events, and this proportion was not balanced, which led us to apply undersampling of the majority class. Therefore, we consider that to use the predictive model in production, it is necessary to train it with a larger dataset that is balanced in both classes of interest.

Considering this second DSR cycle, the focus was on improving the ML model because in the first cycle we adopted a single model as a reference. Therefore, we automatically defined the most appropriate ML model for the dataset of interest, and the results obtained were positive as it was possible to establish an ML model adjusted to the dataset and capable of generating predictions. This model was automatically selected and adjusted to provide adequate performance for the training dataset.

It is worth noting that the model was able to make accurate predictions for a new dataset. However, we highlight that the initial dataset was not optimally prepared for the ML process, and significant effort was dedicated to data preprocessing and statistical data processing for class inference to favor machine learning. Likewise, the volume of data can be increased to present a greater volume of events of the target class to improve the ML process.

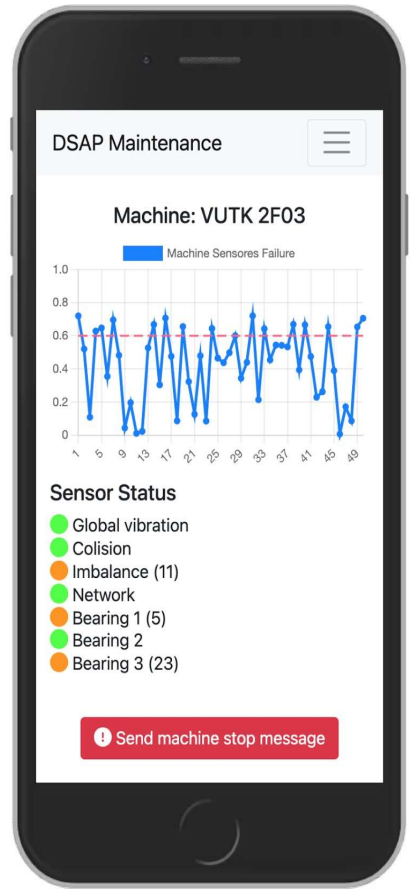

Figure 14 – Mobile Application Dashboard.

Based on ontological processing and specific rules, it was possible to monitor the condition of the machines through graphs from a mobile application (Figure 14). In the mobile application, maintenance managers were able to monitor the evolution of the condition of the machines at hourly intervals, receive notifications about critical failures of each of the sensors, and the status of all machines through the main graph. Figure 14 shows the dashboard of the Mobile Application generated from the DT-Create suite of services.

In Figure 14, it is possible to observe that the operating scenario is configured as moderate, represented by the red dashed line. In this configuration, failures of the machines are tolerated in the interval [0, 0.6]. At times, the sensors report values above the ideal, and the dashboard accumulates these events within the observation window, signaling which sensors presented a failure signal. Thus, given the general condition of the machine and the condition of each of the sensors, it is possible to send a machine stop signal in critical cases. The oscillations in the graph represent the failure signals of the sensors individually, which together represent the machine's current condition.

### 4.5 Results

After Case Study 1 and Case Study 2, we can consider that the results were satisfactory, with some concerns reported in this section. The research question: **"How can decision-making related to predictive maintenance in Industry be supported using DTs?"**, could be answered, considering that in both case studies, a DT was developed to support problems related to predictive maintenance in different industry scenarios. Considering Industry 5.0 support, the human factor was considered, based on the development of user-friendly dashboards with information generated from intelligent processing, whose objective was to facilitate decision-making. The issue of sustainability was also evaluated, considering that the measures taken based on the recommendations generated by intelligent processing allow for a longer useful life for the equipment.

More specifically, we can highlight that the new AutoML component evaluated in the second case study generated a predictive model capable of making predictions with high precision for an unknown dataset (test dataset). However, the initial dataset was not adequately prepared for the machine learning process, and effort was dedicated to pre-processing the data using statistical procedures on the data, which allowed the specification of classes to favor machine learning. Thus, we were able to verify that even with the automatic selection of learning models, considerable time was dedicated to pre-processing, which reinforces that it is not trivial to use new datasets, even with the possibility of selecting learning techniques. Likewise, we highlight the need for a large volume of data to present a larger set of events of the target class to improve the machine learning process.

The Ontology Model component, once instantiated, demonstrated assertiveness in representing the information concerning the failure events indicated by the sensors installed in the machines. Furthermore, this component was able to extract knowledge from the data provided by the AutoML component, enriching the data through appropriate semantic processing.

Thus, the solution proposed in the second DSR cycle, oriented to data on machine operating conditions, and data processing through automated predictive models and semantic models, was able to jointly assess the criticality of failures indicated by sensors associated with semantic processing and the qualitative and quantitative characteristics of the failures, enriching the information. It also produced data that allowed the autonomous agent to act by displaying alert messages on a dashboard to machine operators and managers. Thus, it was possible to indicate that the combined use of ontological techniques and machine learning allowed autonomous software agents to detect equipment failures and maintenance needs.

The two case studies presented distinct contributions to the improvement of the artifact. In the first DSR cycle, the use of ontologies enriched the available data through the processing of ontology inference and context data. In the second DRS cycle, we improved the machine learning process, creating a structure where it is possible to automate, via programming code, selection steps, creation, and distribution adjustment of the ML model. These adjustments are important for the prediction model because there is an agent acting as a service to calibrate the model at time intervals.

Based on the results of Case Study 2, we confronted the theoretical conjectures about the use of ML and ontology after Case Study 1. We analyzed whether, after this interaction, the artifact can support decision-making with the most complete and accurate ML process. The industrial environment where the maintenance process takes place is complex and multifaceted, with many variables and factors that can affect the quality of predictive maintenance. This complexity can make it challenging to develop ML models that can accurately capture the relationships and patterns in the data and make reliable predictions or recommendations.

In addition, ML models for predictive maintenance may need to handle a wide variety of data types and sources, such as sensor data, logs, and other types of data, which can increase the complexity of the model development process. To mitigate this difficulty, we use ontology. Overall, the combination of ontology and ML can significantly improve the insights gained from predictive maintenance systems and enable more data-driven decision-making in maintenance operations.

Regarding the first DSR cycle, the use of neural networks was selected as a reference model for learning on the data, but there was no automated model selection that considered the model's fit with the dataset used in Case Study 1. We improved the artifact in the second DRS cycle, starting with the use of a library for the automatic selection of an ML model. The artifact was also improved in information pre-processing, with automatic outlier removal processes, data normalization, selection of more relevant features, and model quality tests.

As a result, the artifact in the second DRS cycle offers decision support in predictive maintenance processes, either by data enrichment that combines ontologies and context data or by statistical processing and self-selection of learning models.

### 4.6 Threads to Validity

The limitations of this research are mainly related to the quality of the input data. In the first DSR cycle, the data had labels related to the classes of interest (criticality), which facilitated intelligent processing. In the second DSR cycle, the data were used without preprocessing, which may occur in other application scenarios. The absence of preprocessing may influence the interpretation of the results and the ability to generalize the conclusions to other contexts or systems. In the second case study, the lack of preprocessing would invalidate the use of the DT-Create suite of services.

As for **Internal Validity**, we can confirm that models can become overly complex due to overfitting, capturing noise, or random patterns from the training data that do not generalize adequately to certain data. This impairs the ability to infer accurate results. If the data used to train and test the model is not representative of industrial sensor operating conditions, the results may be biased (data selection bias). Also, inconsistencies or errors in the sensor data can lead to erroneous conclusions about the model's performance (measurement bias).

**External Validity** states that the ability to generalize the results to other industrial environments or sensor types may be limited if the dataset used is too specific or not comprehensive. This limitation is known as generalizability. Additionally, using an inappropriate domain ontology may invalidate the results. Similarly, the conditions under which the data were collected may not reflect all possible operating settings, which limits the applicability of the model in different contexts, a problem known as operating settings.

**Construct Validity** is crucial to ensure that a machine learning model accurately represents the problem it is intended to solve. This validation involves two main aspects.

- Representativeness: Models must accurately capture the relationships between sensor data and underlying physical phenomena or industrial processes. Otherwise, models may present inconsistent or irrelevant results.
- Feature suitability: Appropriate feature selection and engineering are essential for models to learn relevant patterns in the data. Poorly chosen features can lead to models that do not correctly represent the problem, resulting in low performance and inaccurate predictions.

Moreover, the predictive validity of models, that is, their ability to predict future results, can be compromised in two situations:

- Difficulty in dealing with unforeseen situations: Models may have been trained on a specific data set and therefore have difficulty predicting results in situations that diverge significantly from the training context.
- Lack of comparison with benchmarks: The evaluation of the performance of the models is incomplete without comparing them with other models or industry benchmarks. This comparison allows us to contextualize the performance of the proposed models and determine whether they represent a significant improvement over existing solutions.

## 5 CONCLUSION

This work established a theoretical framework considering IoT, semantic models, and intelligent decision support systems, identifying from a literature review the challenges related to the use of sensor data for decision-making in maintenance processes. Given the difficulties associated with maintenance decision-making in industry, the work investigated solutions by proposing a suite of services for specifying DTs, with the aim to answer the Research Question:

"**How can decision-making related to predictive maintenance in Industry be supported using DTs?**".

The suite of services, called DT-Create, encompasses a set of components for specifying DTs for collecting, processing, storing, and enriching data obtained through sensors, which can be installed in machines and equipment or even in the production environment. To evaluate the effectiveness of the proposed architecture, two DSR cycles were executed, and two case studies were conducted to provide evidence on the viability of DT-CREATE in specifying DTs for decision-making. Machine learning algorithms were used to analyze data collected and identify patterns and trends that may not be immediately apparent to humans. This analysis can help maintenance managers decide how to manage production assets and improve overall efficiency and mitigate high-critical failures by anticipating maintenance actions. On the other hand, ontologies were used for semantic processing of information generated by machine learning models. In the context of predictive maintenance, ontology helped to organize and classify data collected by sensors, producing alerts about critical scenarios that require intervention.

We combined semantic inferences with machine learning predictions to enrich data and improve decision-making in industrial maintenance. By visualizing predictive results and information about the state of machines on mobile devices, the solution offers managers a relevant tool to increase the efficiency and productivity of operations.

Decisions in predictive maintenance need to consider the diversity of information and devices present in different contexts. In addition, contextual information is generally not used in decision-making due to the complexity of managing heterogeneous data. This work presented a suite of services that aimed to solve the problems of data collection, processing, and visualization to support decision-making.

Our results are relevant because they combine technologies that can drive predictive maintenance in Industry, promoting increased productivity, and competitiveness in industries. Supported by real-time data collection and analysis, the approach supports decisions that prevent or minimize production stoppages due to critical machine failures. The enhancement of the DTs specified through DT-CREATE provides additional benefits, such as:

• Increased equipment availability: Failure prediction allows maintenance interventions to be planned before they occur, maximizing productivity and equipment lifespan.

• Optimization of maintenance resources: Directing the use of labor and spare parts for interventions that are truly necessary generates savings and gains in efficiency.

• Increased operational safety: Minimizing the risk of accidents and unplanned downtime ensures a safer work environment for employees.

The results evaluated were consistent as they address the increasingly important demands of industry. We intend to reinforce and prioritize quality attributes such as flexibility, extensibility, and scalability. We can also define new semantic rules to support the enrichment of data in the ontology and its integration with other domain-specific ontologies to increase the knowledge extraction capacity. New dataset preprocessing techniques can also be used to improve the generalization capacity of DT-Create and its application in several different domains. Finally, it would be helpful to carry out new experiments in other Industry maintenance subdomains to evaluate the support offered by DT-Create in various application domains.

In future work, we intend to generate new instances of DT-Create and integrate them into a software ecosystem. This will allow us to explore collaboration aspects and integration between different business areas, with the aim of improving the decision-making process. Combining and processing additional data sources and sensors can generate more accurate results, reduce costs, and contribute to sustainability in industry.

**Acknowledgements**

This work was partially funded by UFJF/Brazil, CAPES/Brazil, CNPq/Brazil (grant: 307194/2022-1, grant: 313568/2023-5).